\documentclass[aps,prl,twocolumn,superscriptaddress,showpacs]{revtex4}
\usepackage[dvips]{graphicx}
\usepackage{graphicx,color}
\usepackage{amsmath}
\usepackage{bm}
\usepackage{dcolumn}
\begin{document}

\title{Adatoms and clusters of $3d$ transition metals on graphene:

Electronic and magnetic configurations}

\author{T.~Eelbo}
\affiliation{Institute of Applied Physics, University of Hamburg, Jungiusstr. 11, D-20355 Hamburg, Germany}

\author{M.~Wa\'sniowska}
\affiliation{Institute of Applied Physics, University of Hamburg, Jungiusstr. 11, D-20355 Hamburg, Germany}

\author{P.~Thakur}
\altaffiliation[Present address: ]{Diamond Light Source, Rutherford Appleton Laboratory, Didcot OX11 0QX, United Kingdom}
\affiliation{European Synchrotron Radiation Facility, BP 220, F-38043 Grenoble, France}

\author{M.~Gyamfi}
\affiliation{Institute of Applied Physics, University of Hamburg, Jungiusstr. 11, D-20355 Hamburg, Germany}

\author{B.~Sachs}
\affiliation{1$^{st}$ Institute of Theoretical Physics, University of Hamburg, Jungiusstr.~9, D-20355 Hamburg, Germany}

\author{T.~O.~Wehling}
\affiliation{Institut f\"{u}r Theoretische Physik, Universit\"{a}t Bremen, Otto-Hahn-Allee 1, D-28359 Bremen, Germany}
\affiliation{Bremen Center for Computational Materials Science, Am Fallturm 1a, D-28359 Bremen, Germany}

\author{S.~Forti}
\affiliation{Max Planck Institute for Solid State Research, Heisenbergstr. 1, D-70569 Stuttgart, Germany}

\author{U.~Starke}
\affiliation{Max Planck Institute for Solid State Research, Heisenbergstr. 1, D-70569 Stuttgart, Germany}

\author{C.~Tieg}
\affiliation{Helmholtz-Zentrum Berlin, Albert-Einstein-Strasse 15, D-12489 Berlin, Germany}

\author{A.~I.~Lichtenstein}
\affiliation{1$^{st}$ Institute of Theoretical Physics, University of Hamburg, Jungiusstr.~9, D-20355 Hamburg, Germany}

\author{R.~Wiesendanger}
\affiliation{Institute of Applied Physics, University of Hamburg, Jungiusstr. 11, D-20355 Hamburg, Germany}
\date{\today}

\begin{abstract}

We investigate the electronic and magnetic properties of single Fe, Co, and Ni atoms and clusters on monolayer graphene (MLG) on SiC(0001) by means of scanning tunneling microscopy (STM), X-ray absorption spectroscopy, X-ray magnetic circular dichroism (XMCD), and ab-initio calculations. STM reveals different adsorption sites for Ni and Co adatoms. XMCD proves Fe and Co adatoms to be paramagnetic and to exhibit an out-of-plane easy axis in agreement with theory. In contrast, we experimentally find a nonmagnetic ground state for Ni monomers while an increasing cluster size leads to sizeable magnetic moments. These observations are well reproduced by our calculations and reveal the importance of hybridization effects and intraatomic charge transfer for the properties of adatoms and clusters on MLG.

\end{abstract}


\maketitle

In the last decades, the miniaturization of devices has been a main goal in fundamental and applied research. Understanding the smallest magnetic units consisting of a few atoms or even a single atom is a crucial step towards this direction. While magnetic nanostructures on metals~\cite{Gambardella2003, Meier2006}, semiconductors~\cite{Kitchen2006} and insulators~\cite{Hirjibehedin2006} have already been intensively studied, the discovery of graphene allowed to investigate a new substrate with intriguing electronic properties~\cite{Geim2007}. The electronic structure of monolayer graphene (MLG) exhibits a linear energy dispersion with massless Dirac fermions which offers new opportunities for spintronic devices. Tuning the band structure of graphene and at the same time realizing the desired spin textures is a challenging task. For this reason intensive studies on the disorder in graphene introduced by vacancies~\cite{Ugeda2011}, adatoms~\cite{Brar2011}, and various molecules~\cite{Schedin2007} have been performed in the past. On the other hand the nature of the interaction between graphene and individual transition metal (TM) impurities and how the impurities' properties can be modified has been debated mainly in theory (cf. Refs.~\cite{Wehling2011, Chan2008, Cao2010} and references therein) and only a few experiments have been reported so far~\cite{Gyamfi2012, Brar2011}. Especially magnetic properties of single magnetic adsorbates on graphene have experimentally not been investigated so far.

In this article, we study individual 3$d$ TM atoms on graphene prepared by thermal decomposition of the SiC surface in Ar-atmosphere~\cite{Starke2009}. This preparation method results in high quality monolayer graphene, see the supplementary material~\cite{sm} for further details. By means of X-ray magnetic circular dichroism (XMCD) we confirm the presence of sizeable spin and orbital moments and high-spin ground states of Fe and Co adatoms. Instead, Ni adatoms are found to be nonmagnetic but show unexpected X-ray absorption (XAS) spectra. While a vanishing XMCD-signal suggests a $3d^{10}4s^0$ configuration, in line with Ref.~\cite{Gyamfi2012}, XAS shows a multi-peak structure. Thus, XAS indicates a non-fully occupied $d$-shell of the Ni adatoms. We will explain these results based on ab-initio calculations using density functional theory (DFT).

The experiments were performed in two separate vacuum chambers. The scanning tunneling microscopy (STM) measurements were performed inside ultrahigh vacuum using a low-temperature STM operating at 5~K. The XMCD measurements were performed at the ID08 beamline of the European Synchrotron Radiation Facility. Transition metals were deposited in minute amounts, i.e. 0.005~$-$~0.014 monolayer equivalent (MLE), at a substrate temperature of 10~K (in both systems) in order to obtain well-isolated monomers. The oxide free surface during the XMCD study was verified by the absence of a peak at the oxygen $K$ edge in X-ray absorption spectra recorded before and after metal evaporation. XAS at the $L_{3,2}$ edges was performed in the total electron yield mode using circularly polarized light at 10~K. Fields of up to 5~T were applied to align the magnetization at $0^\circ$ (normal) and $70^\circ$ (grazing incident angle) with respect to the surface normal. All spectra have been background corrected. 

To gain a detailed understanding of the XAS of Ni adatoms we performed atomistic simulations of the system using first-principles DFT. Two different packages were used, namely the Vienna ab-initio simulations package (VASP)~\cite{kresse_vasp} with a projector augmented plane wave (PAW) based code~\cite{PAW1,PAW2}, and the Wien2k package~\cite{wien2k} with a full-potential linear augmented plane wave (LAPW)~\cite{andersenLAPW} basis sets. We employed the generalized gradient approximation (GGA)~\cite{GGA1, GGAPBE} to the exchange-correlation potential and converged the results carefully with respect to the number of k-points. The system was modeled by a 3~x~3 graphene supercell of 18 carbon atoms with Ni monomers or Ni dimers placed on top. A vacuum of at least 8~\AA~(Wien2k) and 22~\AA~(VASP) was included. X-ray absorption spectra were calculated within the dipole approximation using Fermi's golden rule. We neglected core-hole effects, but we found in test calculations that these do not significantly change the calculated x-ray spectra.

\begin{figure}
\begin{center}
\includegraphics[width=0.5\textwidth]{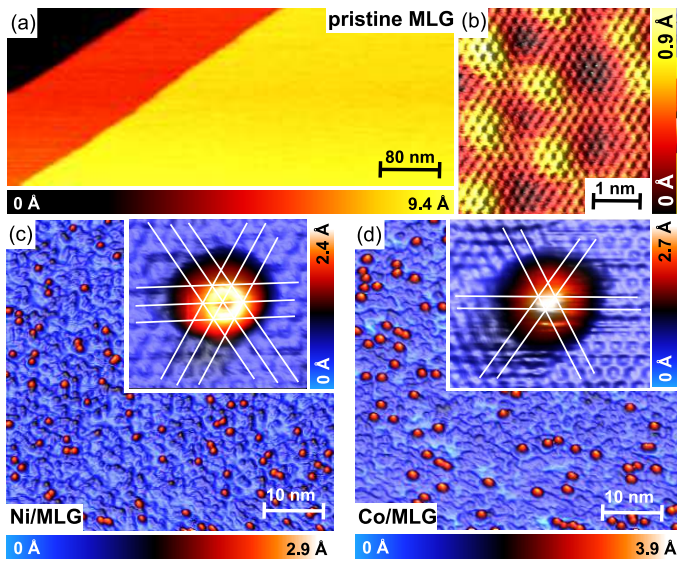}\\
\caption{(color online) (a) and (b): STM topographies of pristine graphene. Tunneling parameters are $U=1$~($-0.1$)~V and $I=0.5$~($0.5$)~nA. (c): Ni on graphene. The inset reveals hollow site adsorption. Tunneling parameters are $U=0.4$~($-0.1$)~V and $I=0.1$~($0.15$)~nA. (d): Co on graphene. The inset exhibits top site adsorption. Tunneling parameters are $U=0.4$~($-0.5$)~V and $I=0.025$~($0.3$)~nA.} \label{Fig1}
\end{center}
\end{figure}

In order to identify the adsorption geometry we performed high-resolution STM measurements, see Fig.~\ref{Fig1}. An overview of MLG prepared on the silicon-terminated SiC(0001) surface before adatom deposition is presented in Fig.~\ref{Fig1}a. A high-resolution image of the surface is shown in Fig.~\ref{Fig1}b exhibiting honeycomb as well as buffer layer corrugation~\cite{Riedl2007}. We note that the surface is fully covered by MLG with a very low defect density which both are required conditions for spatially averaging XMCD measurements in order to avoid buffer/double layer graphene or defect contributions~\cite{sm}. Figures~\ref{Fig1}c and d show constant-current maps of MLG covered by single Ni and Co adatoms, respectively. The insets of subfigures~\ref{Fig1}c and d present the adsorption sites for Ni and Co. In both insets the bright lines are guidelines for the eyes indicating rows of hollow sites. A close inspection shows that in case of Ni the crossing perfectly fits the center of the adatom (hollow site - in line with Ref.~\cite{Gyamfi2012}) while the center of the Co adatom is found to fit a triangle spanned by surrounding crossings (top site). These results perfectly match the predicted adsorption sites found by calculations using a general gradient approximation functional with an on-site Coulomb potential of $U=4$~eV~\cite{Wehling2010, Wehling2011}.

In order to exclude aggregation processes within the XMCD measurements we evaluate the diffusion barriers of the adsorbed Ni, Co, and Fe monomers. From calculated binding energies, we can already extract an expected trend for the diffusion energy among the different TM adatoms on MLG. In a simple approximation the diffusion barrier is given by the difference of the binding energies of different adsorption sites~\cite{Wehling2011}. Using this assumption the diffusion barrier is predicted to be the largest for Ni adatoms. It is especially larger than for Co adatoms which itself is larger than the diffusion barrier of Fe adatoms. Indeed, we revealed by means of the STM experiments Ni adatoms to be most stable in agreement with theory. In case of Ni we find an effective diffusion temperature of $T_{\rm D} \approx 97$~K. Co adatoms are found to be less stable ($T_{\rm D} \approx 50$~K) than Ni adatoms. In case of Fe it was not possible to determine the adsorption site as well as the effective diffusion temperature $T_{\rm D}$ because they were manipulated by the STM tip.

\begin{figure*}
\begin{center}
\includegraphics[width=1\textwidth]{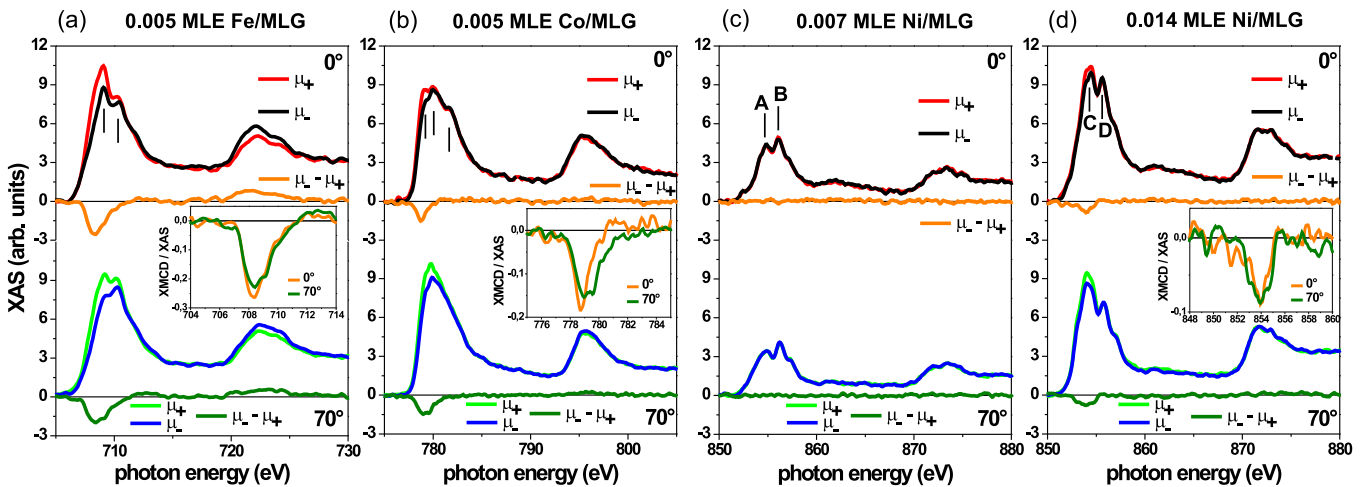}\\
\caption{(color online) XAS and XMCD measurements for (a) Fe (0.005 MLE), (b) Co (0.005 MLE), (c) Ni adatoms (0.007 MLE), and (d) higher Ni coverage (0.014 MLE) at $T=10$~K, $B=5$~T. Upper and lower panels display $L_{3,2}$ edges of XAS and XMCD spectra achieved with parallel ($\mu_{+}$) and antiparallel ($\mu_{-}$) alignment between the helicity of the incident beam and the magnetic field B, obtained at normal ($0^\circ$) and grazing ($70^\circ$) incident angle. The spectra have been normalized with respect to the incident beam intensity and the $L_{3}$ preedge intensity. The insets in subfigures (a), (b), and (d) show the XMCD divided by the averaged XAS intensity for $0^\circ$ and $70^\circ$, respectively.}
\label{Fig2}
\end{center}
\end{figure*}

For exploring the magnetic properties of the adatoms on MLG XAS and XMCD were employed. Fig.~\ref{Fig2}a-c shows the XAS and XMCD spectra of individual Fe, Co, and Ni impurities on graphene. For all impurities the XAS spectra reveal a multi-peak structure at the $L_3$ edge. The observed multi-peaks are contrary to XAS measurements performed on bulk crystals~\cite{Chen1995, Chen1990} and on monomers adsorbed on metallic surfaces (see Refs.~\onlinecite{Gambardella2003, Brune2009, Blonski2010} and references therein) which generally hints toward a weaker hybridization with the substrate in case of MLG as for the aforementioned metal substrates. We note that to our best knowledge XAS/XMCD experiments on TM monomers adsorbed on graphite have not been performed yet. The XMCD signals (see Figs.~\ref{Fig2}a and b) demonstrate the existence of non-zero magnetic moments for Fe and Co while the spin and orbital moments for Ni are zero, cf. Fig.~\ref{Fig2}c. In the following we will first discuss the results for Ni, compare them with ab-initio calculations, and then turn toward Fe and Co.

For a low Ni coverage, most Ni adsorbates are supposed to be monomers.  Naturally, the electronic configuration of free Ni atoms is 3$d^8$4$s^2$. However, on graphene a promotion mechanism of two electrons from the 4$s$ towards the 3$d$-shell fills up the Ni $d$-shell~\cite{Gyamfi2012}. Therefore, we expect Ni monomers adsorbed on MLG/SiC(0001) to show a 3$d^{10}$4$s^0$ configuration in line with previous results~\cite{Gyamfi2012}. For this ground state, on the one hand, no XMCD signal should be observed (see Fig.~\ref{Fig2}c). On the other hand, the XAS spectra are expected to show only step-like features because $2p \rightarrow 3d$ transitions are forbidden and $2p \rightarrow 4s$ transitions would result in step-like features~\cite{Leapman1980}. This expectation relies on the common assumption of a broad 4$s$ band. In contrast, we observe a fine multi-peak structure for Ni adatoms, which usually is observed for non-fully occupied $d$-shells of Ni atoms, as in case of measurements performed for Ni in a $3d^8$ configuration in the gas phase~\cite{Martins2006}.

\begin{figure}
\begin{center}
\includegraphics[width=0.5\textwidth]{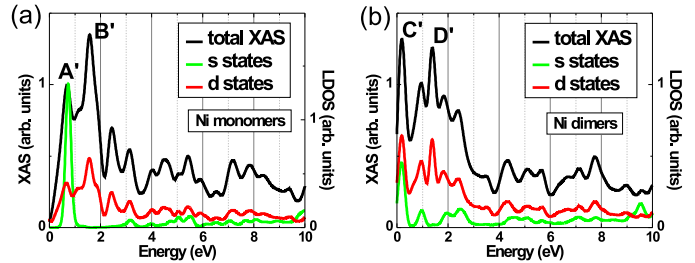}\\
\caption{(color online) Theoretical XAS spectra as well as the $s$- and $d$-LDOS of (a) Ni monomers and (b) Ni dimers on graphene. All spectra and LDOS are spin averaged.} \label{Fig3}
\end{center}
\end{figure}

To understand the multi-peak structure we simulated local density of states (LDOS) and XAS for Ni adatoms. The Ni atoms were placed above the hollow site where the relaxed height is~1.55~\AA. The theoretical monomer spectra (see Fig.~\ref{Fig3}a) exhibit a multi-peak structure at the $L_3$ edge with two pronounced peaks, one at 0.7~eV (A') and a second at 1.6~eV (B'). We attribute these peaks to the two peaks in the experimental spectra at 854.8~eV (A) and 856.1~eV (B), shown in Fig.~\ref{Fig2}c. While the experimental peaks are separated by 1.3~eV, we find 0.9~eV in our simulations. According to Fig.~\ref{Fig3}a the XAS (black) mainly follows the $d$-LDOS (red). Although the empty $s$ orbitals provide the largest peak at 0.7~eV, they show no significant contribution to the XAS because the transition matrix elements for 2$p$ $\rightarrow$ 4$s$ are much smaller than for $2p \rightarrow 3d$ excitations. This is due to the nodal structure of the 4$s$ wave function and the strong localization of the Ni $d$ bands \cite{ebert1996circular}. Hence, the peaks in the monomer case originate almost entirely from unoccupied Ni $d$ orbitals. One possible scenario to explain the origin of the available unoccupied $d$-LDOS is that the electronic configuration of Ni is close to $d^8$ or $d^9$ and exhibits a magnetic moment. However, this is in contradiction to the experimentally determined absence of an XMCD signal and also in disagreement with previous studies \cite{Gyamfi2012}. Our DFT simulations confirm that the Ni monomer exhibits no magnetic moment. Hence, Ni is close to the $d^{10}$ configuration and therefore exhibits a much higher $d$ occupation than bulk Ni. This causes the absence of a local magnetic moment in the Ni adatom: analogous to the case of an Anderson impurity model, there can be no magnetic moments formed close to full occupation. However, a small amount of hole states is still present in the Ni $d$-shell which originates from a hybridization of the Ni $d$-orbitals with the $p_{\rm z}$ states of graphene. The hybridization is anisotropic which leads to the existence of multiple peaks in the LDOS and XAS. Our calculations find that the A' peak in Fig.~\ref{Fig3}a is caused by the hybridization of Ni $d_{{\rm z}^2}$ with Ni $s$ orbitals which marginally contribute to the XAS. In contrast, the B' peak is dominated by a hybridization of $d_{\rm {xy}}$ and $d_{\rm {x}^2-\rm {y}^2}$ states with the graphene $p_{\rm z}$ states. This particular orbital character indeed traces back to the symmetry of the states in the conduction band near the van Hove singularity in graphene, which couple exclusively to this kind of $d$ orbitals (cf. Ref.~\cite{Wehling2010}).

Increasing the coverage of Ni on MLG causes aggregation and therefore leads to clusters on the surface. Again, we find a strong multi-peak feature in the XAS spectra, but most importantly, we now observe a significant XMCD signal at the $L_3$~edge, see the inset in Fig.~\ref{Fig2}d. To explain these findings XAS and LDOS of Ni dimers (with atoms placed on hollow positions; relaxed height:~1.60~\AA) were computed because more adsorbates containing at least two Ni atoms will be formed (cf. Fig.~\ref{Fig3}b). The experimental XAS spectra show two main peaks at 854.4~eV (C) and 855.6~eV (D) (Fig.~\ref{Fig2}d). In agreement we find two main peaks at 0.2~eV (C') and 1.4~eV (D') in our simulated spectra. Hence, the relative intensities between (C) and (D) and the peak-to-peak distance are well-reproduced. Again, XAS mainly follows the Ni $d$-LDOS because the $2p \rightarrow 4s$ transition matrix element is small. The C' peak, corresponding to A' peak in the monomer case, is shifted downwards by $\approx 0.5$~eV. This downshift can be attributed to changes of the $d$-LDOS upon cluster formation. A comparison of the monomer and dimer LDOS shows that the unoccupied dimer $s$~states are shifted towards the Fermi level while more $d$~states above the Fermi level occur. Hence, $s$~orbitals of both Ni atoms hybridize which partially populates $s$~bonding states and depopulates the $d$~orbitals. This effect amplifies for larger clusters. Therefore, sizeable magnetic moments can be obtained for large clusters while Ni monomers are non-magnetic since they are close to a $d^{10}$~configuration. The Ni dimers turned out to be non-magnetic as well -- from additional first-principles calculations we found that Ni clusters with more than four atoms are necessary for a finite magnetic moment. The magnetic properties of Ni clusters, crucially depending on the number of atoms $n$, can thus be tuned between non-magnetic ($n_{\rm {Ni}} \leq 4$) and magnetic ($n_{\rm {Ni}} > 4$) behaviour whereby magnetic moments of about $0.85$~$\mu_{\rm B}$ per Ni atom can be achieved in the limit of infinitely large clusters, i.e. one monolayer of Ni adsorbed on graphene.

In case of Fe we observe a pronounced multi-peak structure in the XAS spectra. The origin of this structure remains unclear and it might be traced back to manybody effects, crystal field, and hybrdization effects. The predicted electronic configuration of Fe monomers on MLG is $3d^74s^1$. The experimentally determined branching ratio ($BR$)~\cite{branching} has a value of $0.74$ and, thus, indicates a high-spin ground state for the Fe monomers in agreement with theoretical predictions~\cite{Wehling2011}. Moreover, Fe has the smallest value for the ratio ($R$)~\cite{ratio} among all studied TM, i.e. $R_{\rm Fe}=0.05\pm0.03$, $R_{\rm Co}=0.35\pm0.02$, $R_{\rm Ni-clusters}=0.42\pm0.02$. Hence, we conclude that Fe atoms exhibit a small orbital moment, in agreement with Ref.~\cite{Sargolzaei2011}. In addition, we determine the easy axis for the magnetic moments from the ratio of the XMCD and the XAS signal strength~\cite{Blonski2010}. For Fe we find this ratio for normal incidence being 20~\% larger than for grazing incidence indicating an out-of-plane easy axis (inset in Fig.~\ref{Fig2}a).

Concerning the XAS measurements for Co the multi-peak structure is less pronounced. We relate this to a stronger hybridization with MLG as compared to Fe. This observation is particularly in line with the predicted binding energies of Co and Fe (0.62~eV and 0.27~$\pm$~0.02~eV, respectively)~\cite{Wehling2011}. The branching ratio of $\approx0.74$, extracted from the XAS spectra, indicates a high-spin ground state for Co as well and is in line with a $3d^84s^1$ electronic configuration of the Co adatoms. The analysis of the relative XMCD intensity of normal and grazing incident reveals an out-of-plane easy axis in agreement with Ref.~\cite{Sargolzaei2011}.

In summary, we find top site adsorption for Co monomers while the adsorption site of Fe adatoms could not be determined due to tip-induced atomic manipulation during our STM measurements. For Fe and Co we observe a multi-peak structure in XAS and a non-zero XMCD signal. Both species are paramagnetic and exhibit an out-of-plane easy axis while the branching ratio indicates high-spin ground states. For single Ni atoms we observe a multi-peak structure in XAS as well while the XMCD shows a zero magnetic moment. Our ab-initio calculations show that this multi-peak structure is due to a small amount of holes in the almost entirely full Ni 3$d$-shell. These holes are due to hybridization of Ni and graphene. Further Ni deposition results in clustering and the rise of non-zero magnetic moments. The main features of the experimental XAS spectra are well reproduced by simulating Ni dimers. Moreover, first-principles calculations reveal a transition from non-magnetic clusters toward magnetic clusters when the Ni cluster size exceeds four atoms, resulting in sizeable magnetic moments.

We thank M.~Sikora, M.~Martins, and M.~I.~Katsnelson for fruitful discussions. Financial support from the ERC Advanced Grant FURORE and the Cluster of Excellence "Nanospintronics" is gratefully acknowledged. This work was supported by the German Research Foundation (DFG) in the framework of the Priority Program 1459 \textit{Graphene} and the SFB 668.


\begin{thebibliography}{31}
\bibitem[{sm()}]{sm}
\eprint{Additional information about the sample quality and uniformity as well as the negligible influence of subsurface SiC step edges onto the properties of the adsorbed TM impurities can be found in the supporting material.}
\bibitem[{rat()}]{ratio}
\eprint{$R$: Ratio of orbital magnetic moment ($m_{\rm L}$) to the sum of spin
  moment ($m_{\rm S}$) and spin dipole moment ($m_{\rm D}$) according to
  $R$~=~$m_{\rm L}$~/~($m_{\rm S}$~+~7~$m_{\rm D}$), where the moments are
  given in $\mu_{\rm B}$/atom and are projected along the incident beam
  direction.}
\end{thebibliography}

\end{document}


\title{Supplementary Material: 

Adatoms and clusters of $3d$ transition metals on graphene: 

Electronic and magnetic configurations}

\author{T.~Eelbo}
\affiliation{Institute of Applied Physics, University of Hamburg, Jungiusstr. 11, D-20355 Hamburg, Germany}

\author{M.~Wa\'sniowska}
\affiliation{Institute of Applied Physics, University of Hamburg, Jungiusstr. 11, D-20355 Hamburg, Germany}

\author{P.~Thakur}
\altaffiliation[Present address: ]{Diamond Light Source, Rutherford Appleton Laboratory, Didcot OX11 0QX, United Kingdom}
\affiliation{European Synchrotron Radiation Facility, BP 220, F-38043 Grenoble, France}

\author{M.~Gyamfi}
\affiliation{Institute of Applied Physics, University of Hamburg, Jungiusstr. 11, D-20355 Hamburg, Germany}

\author{B.~Sachs}
\affiliation{1$^{st}$ Institute of Theoretical Physics, University of Hamburg, Jungiusstr.~9, D-20355 Hamburg, Germany}

\author{T.~O.~Wehling}
\affiliation{Institut f\"{u}r Theoretische Physik, Universit\"{a}t Bremen, Otto-Hahn-Allee 1, D-28359 Bremen, Germany}
\affiliation{Bremen Center for Computational Materials Science, Am Fallturm 1a, D-28359 Bremen, Germany}

\author{S.~Forti}
\affiliation{Max Planck Institute for Solid State Research, Heisenbergstr. 1, D-70569 Stuttgart, Germany}

\author{U.~Starke}
\affiliation{Max Planck Institute for Solid State Research, Heisenbergstr. 1, D-70569 Stuttgart, Germany}

\author{C.~Tieg}
\affiliation{Helmholtz-Zentrum Berlin, Albert-Einstein-Strasse 15, D-12489 Berlin, Germany}

\author{A.~I.~Lichtenstein}
\affiliation{1$^{st}$ Institute of Theoretical Physics, University of Hamburg, Jungiusstr.~9, D-20355 Hamburg, Germany}

\author{R.~Wiesendanger}
\affiliation{Institute of Applied Physics, University of Hamburg, Jungiusstr. 11, D-20355 Hamburg, Germany}
\date{\today}

\maketitle

The particular samples used in this work were prepared by graphitization of the Si-face, the so-called SiC(0001) surface, in Ar-atmosphere~\cite{Starke2009}. As a consequence of this method almost defect free carbon layers can be achieved. Figure~\ref{FigS1}a presents a large scale AFM image of such a sample. The average terrace width of our samples is found to be about $300$~nm. The observable step edges are subsurface steps of the underlying SiC substrate. Figures~\ref{FigS1}b and c show a STM constant current map of a step edge and the respective line profile. The measured apparent height of $2.5$~\AA~matches the spacing of one SiC bilayer. The carbon layer itself is an uninterrupted and closed film continuing across the SiC step edge~\cite{Lauffer2008}. Neither regions of double layer graphene nor regions of an uncovered buffer layer can be found which indicates that the desired coverage of one monolayer of graphene was perfectly achieved.\\

\renewcommand{\thefigure}{S\arabic{figure}}

\begin{figure}[!h]
\begin{center}
\includegraphics[width=0.8\textwidth]{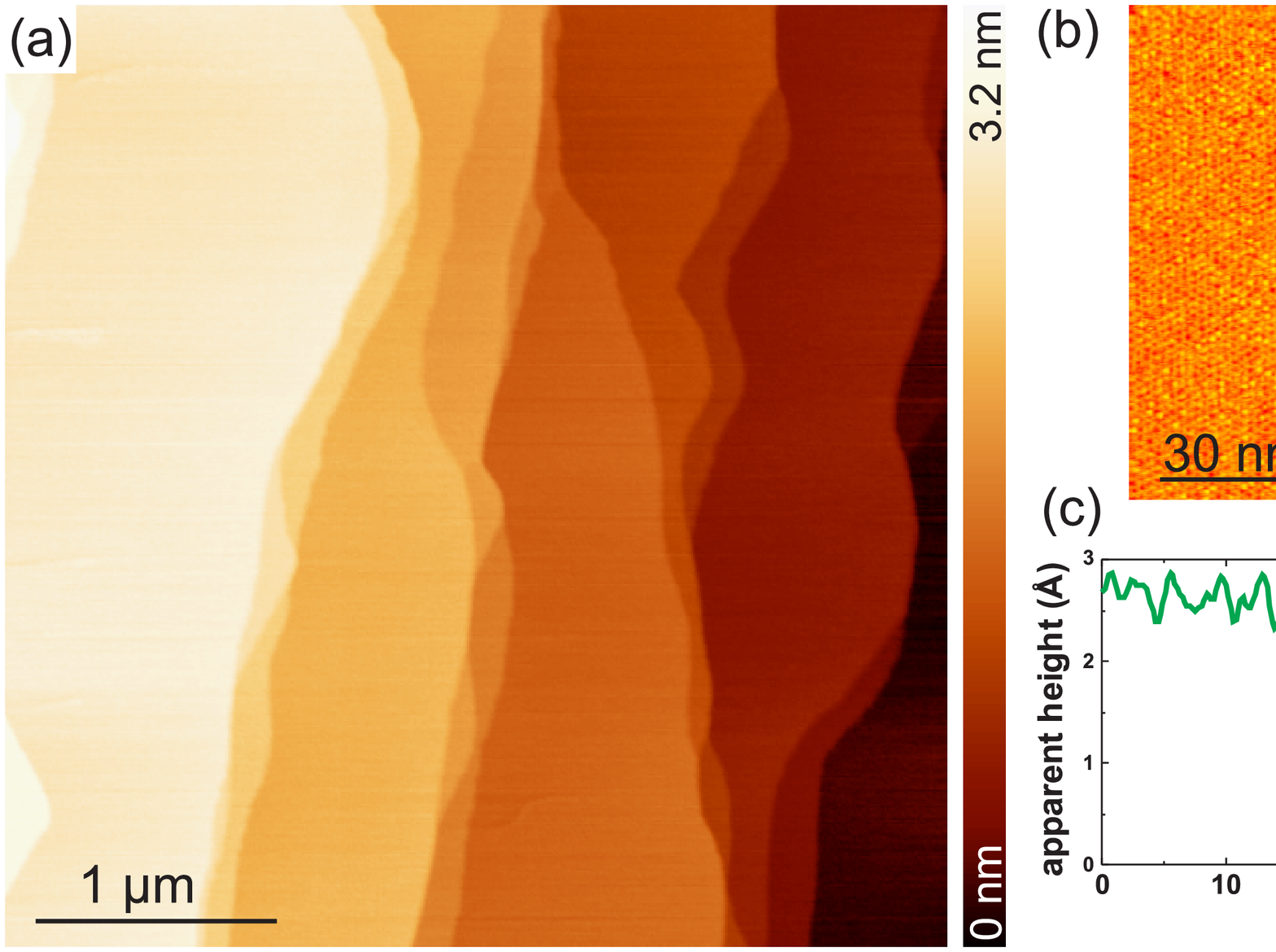}\\
\caption{(color online) (a): Large scale topography of MLG on SiC(0001) measured by means of atomic force microscopy in tapping mode. (b) STM topography of a subsurface SiC step edge indicating the absence of carbon buffer layer or bilayer graphene areas. Tunneling parameters are $U=-1$~V and $I=0.5$~nA (c) Line profile of the SiC step edge as indicated in (b).} \label{FigS1}
\end{center}
\end{figure}

After the low temperature deposition of Ni or Co atoms onto MLG/SiC(0001) we did not find any attractive or repulsive interaction between the adatoms and the SiC step edges by means of scanning tunneling microscopy. The adatoms are statistically distributed across the entire surface. Neither the local density of states (LDOS) of adatoms nor the LDOS of the graphene layer is varying near these subsurface step edges which has been proven by scanning tunneling spectroscopy. Besides the experiments performed directly after low temperature deposition, we heated the samples to variable temperatures above and below the diffusion barrier. We note that single atoms do not interact with the step edges, i.e. the step edges do not change the surface potential of graphene and do not modify the properties of the adatoms. Hence, we exclude step edges to cause multi-peak XAS spectra.

Lauffer and co-workers showed that graphene is bent in the vicinity of an underlying step edge~\cite{Lauffer2008}. This causes an indirect influence on the adatoms, e.g.~the magnetic anisotropy direction of atoms adsorbed in the bent region of graphene is tilted compared to the anisotropy direction of monomers adsorbed on a flat terrace. If we assume an area of $2$~nm (eight unit cells of graphene) in which graphene is not flat, we can estimate the number of adatoms adsorbed in this region from the average terrace width. We expect $0.6$~$\%$ of all adatoms to adsorb on bent graphene, resulting in a negligible influence on the measured XAS and XMCD spectra.